# Inertial Frames and Clock Rates

Subhash Kak

*Abstract:* This article revisits the historiography of the problem of inertial frames. Specifically, the case of the twins in the clock paradox is considered to see that some resolutions implicitly assume inertiality for the non-accelerating twin. If inertial frames are explicitly identified by motion with respect to the large scale structure of the universe, it makes it possible to consider the relative inertiality of different frames.

**Introduction**
According to standard accounts, an inertial frame of reference is one that moves at uniform speed. In Newtonian mechanics, relative to an inertial frame, the motion of a body not subject to forces is always rectilinear and uniform as stated by the law of inertia or the first law of mechanics. Any other frame of reference moving uniformly relative to an inertial frame is also an inertial frame. With respect to an inertial frame, an object or body accelerates only when physical force is applied.

The laws of nature have their simplest form with respect to these frames. In non-inertial frames one needs to introduce fictitious forces, such as the centrifugal force and the Coriolis force in rotating reference frames, which are a consequence of the acceleration of the reference frame itself and not from any physical force acting on the body.

Newton believed that the universe was infinite and homogeneous on a large scale. This assumption is implicit in the identification of inertial frames in classical mechanics. But in current theories the universe is taken to be finite and objects in the universe are accelerating away from each other and, therefore, inertiality is a somewhat different problem. In relativity theory, Lorentz invariance is the invariance of the laws of physics in inertial frames under changes of velocity or orientation**.**

The identification of inertial frames has led to endless debate in the twin or the clock paradox of relativity [1]. If clock rates are related to the velocity of the frame alone with no influence of acceleration, then the two frames are in a symmetric situation. As both twins move with uniform speed (excepting for brief episodes of acceleration and deceleration by one of the twins), the determination as to which of the two is privileged can only be made by considering the speed with respect to the isotropy of the finite universe [2]. If one is prepared to bring gravity into the picture, locally this determination may be made by seeing which twin is associated with the dominant gravitational field. Langevin [1] thought there was no paradox since one of the twins had to accelerate and decelerate in his journey. But this implicitly assumes that the non-accelerating frame is the inertial one.

Einstein's first treatment [3] of the twin paradox, which was in terms of simultaneity and not acceleration, implicitly took the stay-at-home twin to be inertial. When confronted by critiques of this treatment, Einstein later claimed [4] in 1918 that since one of the clocks is in an accelerated frame of reference, the postulates of the special theory of relativity do not apply to it and so "no contradictions in the foundations of the theory can be construed." This later view of Einstein is generally considered to be incorrect [5]. But as we will see, there are others who continue to maintain that gravitation or acceleration is crucial to the resolution to the paradox.

The debate on the twin paradox is not concerning the mathematics of Lorentz transformation but rather the interpretation of the equations. Interpretations within scientific theory are related to the philosophical framework within which the equations of the theory are conceived. For example, the postulates of quantum theory are seen in a variety of frameworks that include the Copenhagen interpretation, the ensemble interpretation, the consistent histories, and the many-worlds view. The elimination of the viability of a specific framework requires the use of careful experiments that have the capacity to distinguish between differing frameworks. For a variety of historical reasons, the matter of interpretation of Lorentz equations has not received the same amount of attention as that of quantum mechanics.

Here we revisit the historiography of the problem of inertial frames. We consider the determination of these frames and argue that certain conditions on the nature of the universe are implicit in the conventions related to the use of Lorentz transformations. We also consider the implication of this identification on time accumulation in different frames.

**Background**
Early models of mathematical astronomy were algebraic (for example [6],[7]) and implicitly assumed a privileged role for the observer. The question of observers with different relative velocity was addressed by the fifth/sixth century astronomer Āryabhaṭa [8] who cryptically suggested that one could only determine the motion of an observer with respect to another observer. Considering the earth to be spinning, he suggested that the experience of the observers was the same at all places including the poles [9]. Galileo in the *Dialogue Concerning the Two Chief World Systems* (1632) argued that one cannot use any mechanical means to determine uniform motion. This idea eventually came to be called the principle of Galilean relativity.

Outside of astronomy, early thinkers considered the question of equivalence of observers in more general and abstract contexts. For example, the Vedic idea of Indra's net (*indrajāla*) models a universe that extends infinitely in all directions. At each node of the net where threads cross there is a gem that mirrors all the other gems in the net. The reflection of each gem in every other one represents the idea that each point of reality is defined by the rest of the universe. In the Brahmavaivarta Purāṇa (4.47), in the chapter on Indra and the Ants, it is argued that there are countless number of suns and planets. In the *Vijñāna Bhairava* 133 it is stated that reality exists in Indra's net without materiality (अतत्त्वम् इन्द्रजालाभम् इदं सर्वमवस्थितम्, *atattvam indrajālābham idaṃ sarvamavasthitam*). In other words, it is the relationships between the parts that define the whole and materiality itself is a manifestation of such relationships.

The threads of Indra's net are the laws that bind the universe. There are other similar threads in our mind that make it possible for observers to comprehend reality [10],[11]. In these conceptions it is taken that time flows at different rates for different observers but these are considered in abstract planes and not kinematically.

Newton considered space to be absolute although he recognized that there exist "relative spaces" in any of which true forces and masses, accelerations and rotations, have the same objectively measured values. The normal understanding in physics is that in an inertial frame free particles move in straight lines, although in general relativity rectilinear motion is replaced by motion along geodesic lines. But for any trajectory one can define some coordinate system in which it is rectilinear and thus we cannot identify an inertial system by the motion of a single particle. For an inertial system at least three non-collinear free particles should move in non-coplanar straight lines. Ernst Mach claimed that the law of inertia and Newton's laws generally implicitly appeal to the fixed stars as a spatial reference-frame.

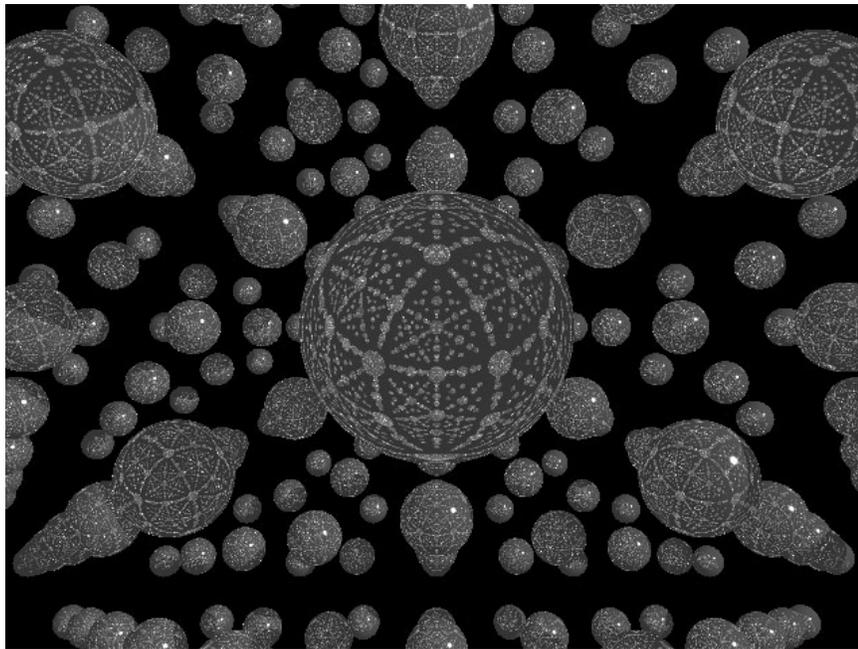

An artistic visualization of Indra's net using computer fractals (I.N. Galidakis)

A few years after the Michelson-Morley experiments showed that there was no need to postulate a medium for the propagation of light, Poincaré enunciated the principle of relativity to mean "the laws of physical phenomena should be the same, whether for an observer fixed, or for an observer carried along in a uniform movement of translation; so that we have not and could not have any means of discerning whether or not we are carried along in such a motion" [12]. In this view, all uniformly moving frames are inertial [13].

This raises two basic problems: first, why are inertial frames privileged; second, how to determine if a given frame is inertial. These problems were stated by Einstein as follows:

> What justifies us in dispensing with the preference for inertial systems over all other co-ordinate systems, a preference that seems so securely established by experiment based upon the principle of inertia? The weakness of the principle of inertia lies in this, that it involves an argument in a circle: a mass moves without acceleration if it is sufficiently far from other bodies; we know that it is sufficiently far from other bodies only by the fact that it moves without acceleration. Are there, in general, any inertial systems for very extended portions of the space-time continuum, or, indeed, for the whole universe? [14; page 62]

Poincaré showed that unique synchronization of clocks by means of light signals was possible in every inertial reference system defined by uniform motion alone [15] but he did not consider the question of identification of inertial frames.

One way to determine if a frame is inertial is to compare its time rate to that of all others: if the time accumulated on it is more than on any other, then the frame is inertial, assuming of course that the effects of gravity are identical in both cases or can be factored in the comparison. Since clocks in a stronger gravitational field move at a slower rate with respect to a clock at, say, the north pole, a clock at the equator will move faster because of its greater distance from Earth's center of mass due to the flattening of the Earth, but slower because of its relative speed due to Earth's spin and these two effects cancel each other out on the surface of the Earth. Such a comparison of time accumulations is made in special cases as in the on-board atomic clocks on the Global Positioning System (GPS) satellites.

If we cannot use time accumulation, we must use some other global information to identify inertial frames. Contrariwise, we must show that local information suffices to identify inertial frames but as the remarks of Einstein quoted above show, there is no evidence in support of such a supposition.

**Inertial Frames in the Twin Paradox**

In the twin paradox, Twin A moves inertially from event 1 to event 2, and Twin B undergoes a period of acceleration and then moves inertially from event 1 to an intermediate event 3, where he changes his state of motion, and then moves again inertially from 3 to 2. It is claimed that the total elapsed time of Twin A exceeds that of Twin B.

The paradox is due to the fact that on the one hand Lorentz transformations are logically consistent and on the other both twins are almost entirely in "inertial" motion and if the situation is nearly symmetrical Twin B should not experience less elapsed time than Twin A. Indeed, as pointed out by Max von Laue, one can imagine Twin B to hop on to an inertial frame and then hop back on another inertial frame to return to Twin A.

There is no clear way to distinguish inertial worldlines from all possible worldlines excepting by determining which has experienced more elapsed time which is a circular definition. It is for this reason that Einstein said [16]: "In classical mechanics, and no less in the special theory of relativity, there is an inherent epistemological defect which was, perhaps for the first time, clearly pointed out by Ernst Mach." He further suggested in the same paper that distant masses which we have not included in the system under consideration should be used in determination of inertial motion. Later, Einstein loosened his embrace of Mach's ideas in the development of the general theory of relativity.

In 1918, Einstein suggested that gravitational fields are necessary for resolution of the twin paradox:

> However, while according to the special theory of relativity a part of space without matter and without electromagnetic field seems to be characterized as absolutely empty, e. g. not characterized by any physical quantities, empty space in this sense has according to the general theory of relativity physical qualities which are mathematically characterized by the components of the gravitational potential, that determine the metrical behavior of this part of space as well as its gravitational field. [17]

Born agreed with this view saying: [18] "the clock paradox is due to a false application of the special theory of relativity, namely, to a case in which the methods of the general theory should be applied." This view was also supported by Pauli [19]. On the other hand, Taylor and Wheeler [20] assert that general theory should not be needed, and the invocation of gravitational fields is no longer popular in the analysis of the twin paradox. Yet, contradictory explanations continue to be provided for the resolution of the paradox. There are those who say Lorentz equations suffice to resolve the problem [1] and others who claim that one must invoke acceleration and deceleration [21].

In textbook accounts it is assumed that the twins belonged to an inertial frame before Twin B took off on his rocket, and now, using worldlines on a Minkowski diagram it is shown that Twin B's elapsed time is less. *This account implicitly assumes that any frame that ends up with uniform velocity different from that of the original frame is no longer inertial.*

But this means that uniform velocity is not sufficient in itself in deciding whether a frame is inertial. We need to go back in the history of the frame to see how it came to be endowed with its velocity. *Only frames that have never experienced acceleration in their remote past are inertial!* Inertial frames can thus be identified by considering their place in the cosmological model associated with the universe.

This brings one back to the problem of identification of inertial frames that was mentioned by Einstein. In an infinite, homogeneous universe, there are an infinite number of inertial frames which may be identified by experiment. Likewise, in a universe with a unique origin as in Big Bang cosmology, the identification of inertial frames should also be based on information that is potentially available locally.

Another perspective on the inertial frame problem is that logical paradoxes arise both in formal systems and in the consideration of the entire universe [22],[23] and, therefore, the matter cannot be completely resolved using analysis alone. Some logical paradoxes may be eliminated by enlarging the domain of the discourse and by the design of new experiments that are able to distinguish between different interpretive frameworks.

**Violations of Lorentz Symmetry and General Relativity**
Investigation of the violations of Lorentz invariance has become an important research area in the context of quantum gravity, extensions to the Standard Model of particle physics, and cosmology [24],[25]. For example, the breaking of Lorentz symmetry leads to a breaking of the CPT symmetry which is required in quantum gravity. Likewise, the interaction of quintessence with matter requires that Lorentz symmetry be broken. But these are issues beyond the scope of this paper.

When rotational and accelerated motion is taken to be relative, the twin paradox has new, surprising features [26]. Thus there can be instances when the accelerating twin is older [27].

**Discussion**
The essay has presented a brief history of the problem of inertial frames. The twin paradox of relativity is about the problem of defining inertiality.

Infinite regress is associated with inertiality if only simultaneity is used for the identification of inertial frames. Asserting Twin B is not inertial because it experienced acceleration at some point in its history compels an examination of the history of each frame from the very beginning. But such previous history of the frames is unknowable in general and, therefore, the traditional approach cannot resolve the issue.

The idea that an inertial frame is known merely by its uniform velocity may suffice in a local homogeneous area in a larger universe with a more complex structure. We could also speak of frames that are *approximately* inertial related to a gravitational field. For approximate frames that are undergoing acceleration and moving with different speeds with respect to the rest of the universe, the extent of the departure will determine which frame will have lower time accumulation.

**Acknowledgement.** I thank Nick Percival and Gurpur Prabhu for comments and discussions. I am also thankful to I.N. Galidakis for permission to use his computer fractal visualization of Indra's net.

**References**
1. P. Langevin, L'Évolution de l'espace et du temps. Scientia 10: 31–54, 1911.
2. S. Kak, Moving observers in an isotropic universe. International Journal of Theoretical Physics 46: 1424-1430, 2007.
3. A. Einstein, Zur Elektrodynamik bewegter Körper. Annalen der Physik 17: 891–921, 1905.
4. A. Einstein, "Dialog about objections against the theory of relativity." Die Naturwissenschaften 48: 697-702, 1918.
5. P. Pesic, Einstein and the twin paradox. Eur. J. Phys. 24: 585-590, 2003.
6. S. Kak, Astronomy of the Vedic altars. Vistas in Astronomy 36: 117-140, 1993.
7. S. Kak, The astronomy of the age of geometric altars. Quarterly Journal of the Royal Astronomical Society 36: 385-396, 1995.
8. K.S. Shukla and D.V. Sarma, Āryabhaṭīya of Āryabhaṭa. Indian National Science Academy, 1976.

9. R.H. Narayan, Are Aryabhata's and Galilean relativity equivalent? http://arxiv.org/abs/0802.3884
10. S. Kak, Active agents, intelligence, and quantum computing. Information Sciences 128: 1-17 2000.
11. S. Kak, The universe, quantum physics, and consciousness. Journal of Cosmology 3: 500-510, 2009.
12. H. Poincaré, L'état actuel et l'avenir de la physique mathématique. Bulletin des sciences mathématiques 28 (2): 302-324, 1904.
13. E.T. Whittaker, A History of the Theories of Aether and Electricity. Nelson, 1953.
14. A. Einstein, The Meaning of Relativity. Princeton University Press, 1923.
15. A.A. Logunov, Henri Poincaré and Relativity Theory. Nauka, 2005.
16. A. Einstein, The foundation of the general theory of relativity. Annalen der Physik 49, 1916.
17. A. Einstein, Dialog über Einwände gegen die Relativitätstheorie, Die Naturwissenschaften, 29 November 1918.
18. M. Born, Einstein's Theory of Relativity. Dover, 1962.
19. W. Pauli, Theory of Relativity. Dover, 1981.
20. E.F. Taylor and J.A. Wheeler, Spacetime Physics. W.H. Freeman, San Francisco, 1992.
21. E. Minguzzi, Differential aging from acceleration, an explicit formula. Am. J. Phys. 73: 876-880, 2005.
22. S. Kak, The Nature of Physical Reality. Peter Lang, 1986.
23. S. Kak, Hidden order and the origin of complex structures. In Origin(s) of Design in Nature. Swan, L.S., Gordon, R., and Seckbach, J. (editors). Dordrecht: Springer, 2012.
24. M. Pospelov and M. Romalis, Lorentz invariance on trial. Physics Today 57(7): 40-46, 2004.
25. A. Kostelecky and N. Russell, Data table for Lorentz and CPT violation. Rev. Mod. Phys. 83: 11-31, 2011.
26. Ø. Grøn, The twin paradox and the principle of relativity. arXiv:1002.4154
27. M.A. Abramowicz and S. Bajtlik, Adding to the paradox: the accelerated twin is older. arXiv: 0905.2428.